\documentclass{article}

\usepackage{arxiv}

\usepackage[utf8]{inputenc} 
\usepackage[T1]{fontenc}    
\usepackage{hyperref}       
\usepackage{url}            
\usepackage{booktabs}       
\usepackage{amsfonts}       
\usepackage{bm}
\usepackage{amsmath}
\usepackage{nicefrac}       
\usepackage{microtype}      
\usepackage{lipsum}

\usepackage{graphicx}
\usepackage{float}
\usepackage{amsmath}

\usepackage{subcaption}
\usepackage{array}
\usepackage{multirow}
\graphicspath{ {./figures/}, {./figures/success_segs/}}

\title{Weakly supervised training of pixel resolution segmentation models on whole slide images}

\author{
  Nicolas Pinchaud\\
  \texttt{nicolas.pinchaud@contextvision.se} \\
}

\begin{document}
\maketitle

\begin{abstract}
We present a novel approach to train pixel resolution segmentation models on whole slide images in a weakly supervised setup. The model is trained to classify patches extracted from slides. This leads the training to be made under noisy labeled data. We solve the problem with two complementary strategies. First, the patches are sampled online using the model's knowledge by focusing on regions where the model's confidence is higher. Second, we propose an extension of the KL divergence that is robust to noisy labels. Our preliminary experiment on CAMELYON 16 data set show promising results. The model can successfully segment tumor areas with strong morphological consistency.
\end{abstract}


\section{Introduction}
Pathologists generally base their diagnosis on the presence of localized morphological bio-markers in histopathological images. Training deep neural networks on Whole Slide Images (WSIs) for automatic disease detection and localization is the main approach for the development of decision support tools. However it requires large annotated datasets that are difficult to construct because annotations at the pixel resolution are often received through a costly, noisy, and low bit-rate channel. Weakly supervised approaches aim to tackle the problem of data acquisition by relying on a supervision available on the WSI level. For cancer detection task, the WSIs are labeled as $benign$ or $malign$ depending on the cancer presence. This information is cheaper to acquire, more robust, and allows construction of larger datasets. A model is trained using this high level labeling to retrieve localized, pixel level, disease information.

We propose to use a segmentation model that produces pixel level unnormalized log evidences (or logits) that represent localized cancer presence probabilities in WSIs. We form WSI level cancer presence probability from aggregation of pixel level logits. We can train the model to maximise the mutual information between this probability and the WSIs label. Our assumption is that this serves as a proxy to maximize the pixel level mutual information between pixel wise logits and the (unknown) cancer localization.

However this training scheme requires the segmentation model to work on all the pixels of a WSI at a time. Since a WSI often contain an order of $10^9$ pixels, that makes this approach intractable. To tackle this problem, we propose to train the model on patches randomly sampled from the WSIs. A patch is given a label that is inherited from the slide it is sampled from to form training samples. The drawback is that this can lead patches to be incorrectly labeled. For instance, the cancerous region of a malignant WSI can represent less than $0.1\%$ of the tissue. Therefore, there is high chance to sample patches from benign regions, they will be incorrectly labeled as malignant. This produces noisy supervision that would prevent the model to converge. We propose to address this with two things. First, we propose an extension of the Kullback–Leibler (KL) divergence that is robust to noisy labels.
Second, we sample patches dynamically during training. The sampling is made using a distribution defined by the pixel wise cancer probability map given by the model. Patches are sampled more frequently from regions were the model gives higher cancer probability. In the beginning of the training, the sampling distribution is uniform over the whole slide. As the model converges, the patches are more often sampled from regions containing cancer, thus reducing labeling noise and further improving convergence.

We demonstrate our approach on the CAMELYON 16 dataset \cite{cam16} and show that the model is able to segment tumors at pixel resolution. Even more, the segmentations are strongly correlated with the morphological structures of the tissues. We evaluated the ability of the model to classify slides as malign or benign. The model reached a ROC AUC score of $0.82$ which is a promising score given that only one set of hyper-parameters have been tested.

\section{Related work}

Image segmentation at pixel-level under weak supervision has been studied in \cite{a,1411.6228,1803.07703}. These approaches propose to work under the Multiple Instance Learning framework (MIL). The images are seen as bag of pixels or patches. The elements of the bags have an underlying label that is not available. However a label at the bag level is available and relate to the label of its elements. A negative (i.e. $benign$) bag contains only negative elements, while a positive (i.e. $malign$) bag contains at least one positive element. A convolutional neural network (CNN) is used to produce pixel (patch) level features that are aggregated through a pooling function to form the bag level class prediction. However these approaches need the image to fit into memory and do not scale on WSIs that contain giga pixels.

Weakly supervised approaches on WSIs divide the slide into a grid of tiles \cite{1805.06983, 1802.02212, 1701.00794}. Each tile is processed through a CNN to produce a score that allows them to be ranked. The top/bottom tiles are then used to train a WSI classifier. The gradients are propagated into the scoring CNN up to some depth allowing to learn the scores. The scores of tiles are then interpreted as disease localization confidences. These approaches have several constraints. First, they produce disease location at tile resolution and not at pixel resolution. Second, they are limited in their ability to train complex models. In \cite{1802.02212} the tiles are, prior to the training, mapped into a fixed feature space using a deep neural network trained on ImageNet \cite{imagenet_cvpr09}. A simpler parameterized function is trained over this feature space to produce the scores, consequently limiting the representation power of the learned model. In \cite{1805.06983}, the scoring CNN is applied on all the tiles at the beginning of each epochs, the training is made only on the top tile of each slide thereafter. This leads to slow convergence that requires larger dataset.

In \cite{streamsgd} the authors propose to scale the training of a CNN on WSIs using the streaming stochastic gradient descent method \cite{1804.05712}. This approach does not allow to train segmentation model architectures because it requires upper feature maps layers to fit in memory which prevents the usage of up-convolution layers. The authors propose to extract a saliency map using the gradients at the input pixels.

In \cite{Momeni438341} the authors propose the use of a deep recurrent attention model that classifies a WSI using information provided from a limited number of patches (or glimpses) using visual attention method. Disease localisation is inferred from the glimpses location.

Our approach combines the ability the train pixel resolution disease segmentation models on WSIs without constraining complexity of the learned CNN. Moreover, we do not train on pre-tiled slides, instead we learn on patches dynamically extracted during training according to the current knowledge state of the model.

\section{Model}
\label{sec:headings}
Our model is defined using a segmentation model represented by a parameterized function $M_\theta: [0,1]^{h\times w \times c} \rightarrow \mathbb{R}^{h\times w \times 2}$. $M_\theta$ can be an instance of any kind of state-of-the-art segmentation model such as U-net\cite{ronneberger2015} or DeepLab\cite{deeplabv3plus2018}. For any given image $X$ of dimension $h\times w$ with $c$ channels, the model gives $L:= M_\theta(X) \in \mathbb{R}^{h\times w \times 2}$ representing the un-normalized log probabilities (logits) for benign and malign classes for each pixels. We get the classes probabilities of a pixel $X_{i,j,.}$ using the softmax function:
 $$Q_{i,j,k} = \frac{e^{L_{i,j,k}}}{e^{L_{i,j,0}}+ e^{L_{i,j,1}}}$$
 
 In a weakly supervised setup the pixel level annotation is missing. Instead, the supervision is available on the WSI level. A WSI can either be benign or malign. A benign WSI have all its pixels belonging to the benign class while a malign WSI has at least one of its pixel being malignant. Hence we can translate the segmentation problem to a constrained WSI binary classification problem. The model outputs the WSI's class by computing the slide level benign and malign logits (we note respectively $l_0$ and $l_1$) with:
 $$\textit{l}_{0} = \mathbb{E}[L_{.,.,0}]$$
 
 A WSI is malignant if at least one of its pixels is malignant. We can translate that relationship with the following expression of $l_1$:
$$\textit{l}_{1} = max(L_{.,.,1})$$
 
 However, such strategy results into slow and noisy learning of the model because the gradients would flow through only one pixel at a time. Therefore, we generalize to a \textit{top K} expression:
$$\textit{l}^\eta_{1} = \mathbb{E}[L_{.,.,1}|L_{.,.,1}>P_n]$$
where $P_\eta$ is the $100-\eta$ percentile. For low $n$ we retrieve the max definition of $\textit{l}_{1}$ and for $\eta=100$ we get the same average expression of $l_0$. As for the pixels probabilities, we compute the WSI's classes probabilities using the softmax function. We note $q_0$ and $q_1$ the probability of the WSI to be respectively benign and malignant.

If the dimensions $h \times w$ are not too large, the model can be trained using the cross-entropy loss function on the WSI level probabilities. However, in general $h \times w$ is large and the training is intractable.

\subsection{Training on patches using dynamic sampling}

Instead of training on the full WSI which is intractable, we propose to train on dynamically extracted patches. The model is trained by classifying the patches instead of the WSIs. The patches inherit the class of the slide they are extracted from. In that setup a patch extracted from a benign slide have all its pixels being benign. However, a patch extracted from a malign slide have high chance to contain no malign pixels. This breaks the assumption that at least one pixel have to be malign within $X$ and makes the patch level supervision noisy. We propose to tackle this problem by using an extended version of the KL divergence loss that is robust to noise:

    $$\mathcal{L}_{\bm{\beta}}(\mathbf{p} \| \mathbf{q}) = -\sum_i \frac{p_i}{\boldmath{\beta_i}}\left[\left(\frac{q_i}{p_i}\right)^{\mathbf{\beta_i}} -1\right]$$
where $\mathbf{p}$ is the true distribution and $\mathbf{q}$ the model's distribution. Since:

\[ \lim_{\beta \to 0} \frac{p}{\boldmath{\beta}}\left[\left(\frac{q}{p}\right)^{\mathbf{\beta}} -1\right] = p \log\left(\frac{q}{p}\right) \]

we can properly extend $\mathcal{L}_{\bm{\beta}}$ for the terms where $\beta_i=0$, and particularly we have $\mathcal{L}_{\bm{0}}(\mathbf{p} \| \mathbf{q}) = D_{KL}(\mathbf{p} \| \mathbf{q})$.

If $\forall i, \beta_i<1$, then we have  $\mathcal{L}_{\bm{\beta}}(\mathbf{p} \| \mathbf{q}) =0$ iff $\mathbf{p} = \mathbf{q}$. The vector $\bm{\beta}$ allows to control the shape of the loss for the different classes.  

\subsubsection{Tuning $\beta$ under noisy supervision}

Let $\gamma$ be the probability that a patch $X$ from a malign slide does not contain any malign pixel. These patches are indistinguishable from patches extracted from benign slides and will confuse the model. Since they contain no malign pixel they should be classified as benign. We can leverage the noise induced by the label inconsistency by adjusting $\bm{\beta}$.

Let's suppose we have a data-set made of $malign$ and $benign$ slides. Let's suppose that we can extract one pixel patches from a slide such that $X \in \{0, 1\}$ where $X=0$ is a benign pixel such that we have $P(X=0|malign)=\gamma$ and $P(X=0|benign)=1$.
Let's define a trivial model $\Psi_{\bm{\theta}}$ with $\bm{\theta} = \begin{pmatrix} \theta_0\\ \theta_1\end{pmatrix} \in [0,1]^2$, such that $P(malign|X=0, \bm{\theta}) = \theta_0$ and $P(malign|X=1, \bm{\theta})= \theta_1$. Ideally, we would like the model to classify the benign pixel as benign with probability one, i.e. we would like $P(malign|X=0, \bm{\theta}) = \theta_0=0$.

Optimization of this model using $\mathcal{L}_{\bm{\beta}}$ would give the following gradient over $\theta_0$: $$\frac{\partial \mathbb{E}_X[\mathcal{L}_{\bm{\beta}}(\mathbf{p} \| \mathbf{q})]}{\partial \theta_0} = (1-r)\gamma\theta_0^{\beta_1-1}-r(1-\theta_0)^{\beta_0-1}$$

where $r = P(benign)$ is the ratio of benign slides.

In the case of the KL divergence, obtained with $\bm{\beta}=\bm{0}$, under a balanced data-set $r=\frac{1}{2}$, the minimum would be reached on $\theta_0 = \frac{1}{1+\frac{1}{\gamma}}$. For instance, a noise of $\gamma = 33\%$ would give $\theta_0\approx 0.25$.

We can leverage the noise and obtain lower value of $\theta_0$ by tuning $\bm{\beta}$. The minimum on $\theta_0$ is obtained when $\frac{\partial \mathbb{E}_X[\mathcal{L}_{\bm{\beta}}(\mathbf{p} \| \mathbf{q})]}{\partial \theta_0}=0$ which gives the following value of $\beta_1$:
$$\beta_1  = \frac{1}{\log \theta_0}\left[(\beta_0-1)\log (1-\theta_0) - \log \left(\frac{(1-r)\gamma}{r}\right)\right]+1$$

\begin{figure}
  \centering
  \includegraphics[width=0.49\textwidth,scale=1.0]{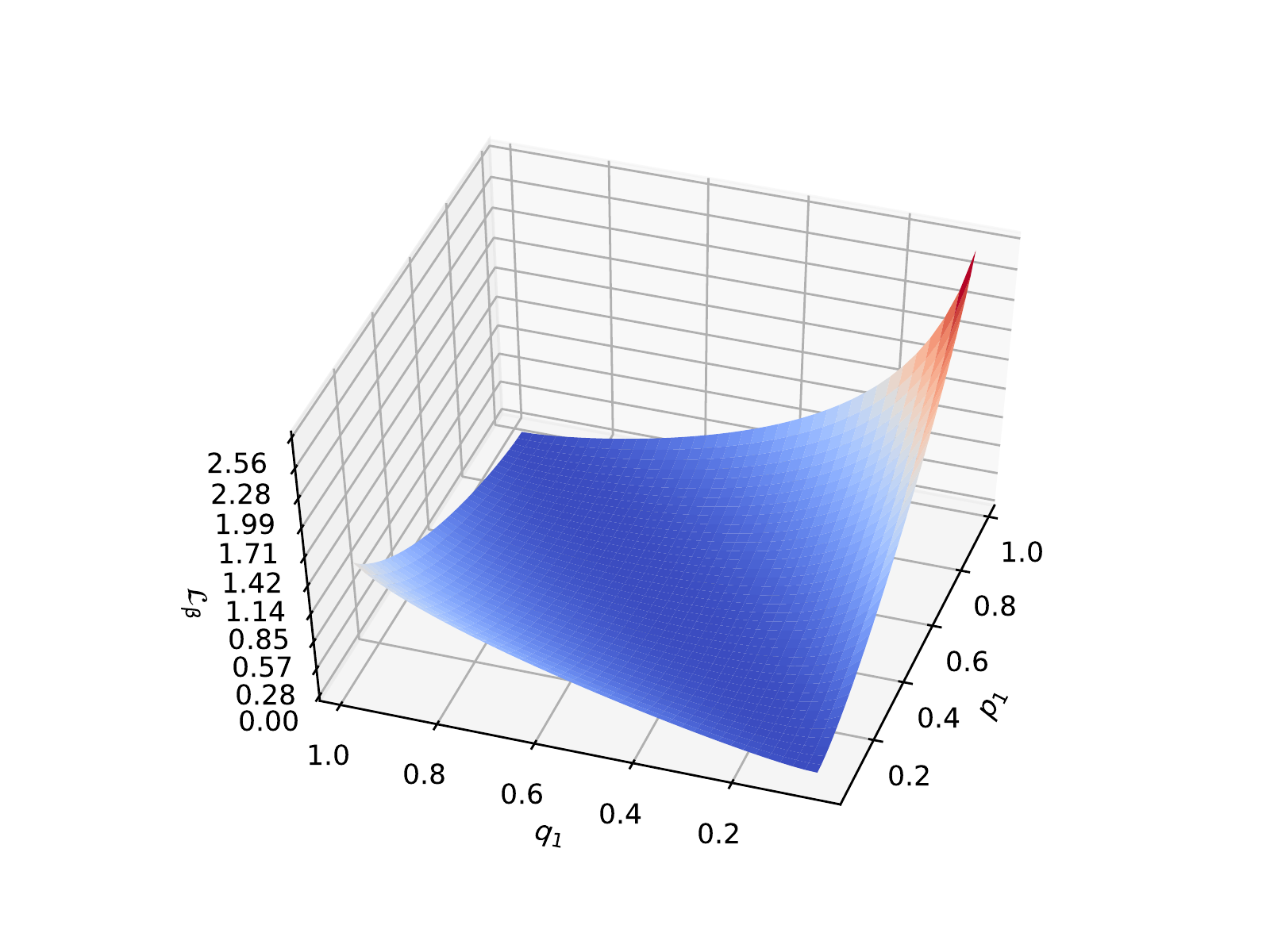}
  \includegraphics[width=0.49\textwidth,scale=1.0]{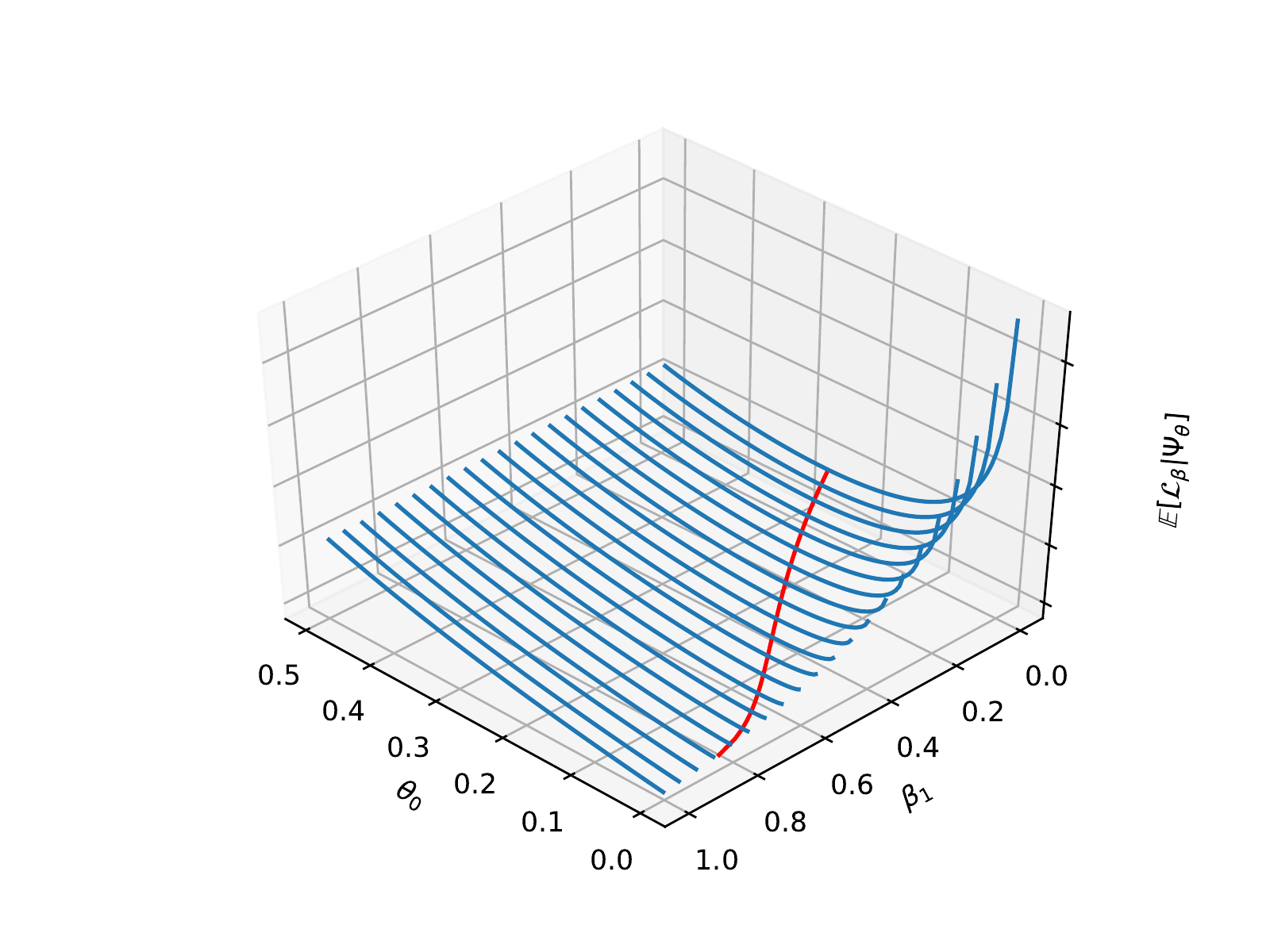}
  \caption{\textbf{left}: The divergence $\mathcal{L}_{\bm{\beta}}$ between two Bernoulli distributions with parameters $p_1$ and $q_1$, with $\beta_{0}=0$ and $\beta_{1}=0.61$. The figure shows the asymmetry induced by $\bm{\beta}$. The loss penalizes differently the false positives and the false negatives. \textbf{right:} Loss of the model $\Psi_{\bm{\theta}}$ given $\theta_{0}$ and $\beta_{1}$ under a noise of $\gamma=33\%$ and with $\beta_{0}=0$. The red line shows the optimal $\theta_{0}$ for a given $\beta_{1}$. We can see that higher values of $\beta_{1}$ allow better mitigation of the noise with lower values of optimal $\theta_{0}$.}
  \label{fig:loss_beta}
\end{figure}

If we have $r=\frac{1}{2}$ and set $\beta_0=0$, if $\gamma=33\%$ and if we wish to get $\theta_0=5\%$, we can set $\beta_1 \approx 0.61$.
The figure \ref{fig:loss_beta} shows how the parameter $\bm{\beta}$ shapes the loss for two Bernoulli distributions. The false positives are less penalized than the false negatives. This affects the gradient by reducing the slope of the loss for the miss-classified examples. Note that our approach is different than scaling the gradients differently for each class. In fact, with this approach, also called \textit{balanced cross entropy} in \cite{focal_loss}, the gradient ratio between a well classified and a miss-classified example would remain constant. In our approach this ratio depends on the parameter $\bm{\beta}$. Also, we differ from the focal loss \cite{focal_loss} as this approach is designed to down weight well classified examples to focalize more on the miss-classified ones. This relies on the assumption that the label information is not noisy.

\subsubsection{Training with dynamic sampling of patches}

We train $M_\theta$ by sampling patches from slides according to a distribution defined by the probability map of the model on the slides. Given a WSI $S \in [0,1]^{h\times w \times c}$, the probability that the pixel $S_{i,j,.}$ is $malign$ according to the model is $Q_{i,j,1}$. The probability to sample a patch centered on the pixel $(i,j)$ is given by:
$$P(i,j|S,\theta,\alpha)= \frac{Q_{i,j,1}^\alpha}{\sum_{u,v}Q_{u,v,1}^\alpha}$$
where $\alpha\geq 0$ controls the entropy of the distribution. With $\alpha=0$, the distribution is uniform and all pixels are sampled with equal probability, while higher values narrow the distribution on pixels having higher classification confidence.

\begin{figure}
  \centering
  \includegraphics[width=0.6\textwidth,scale=1.0]{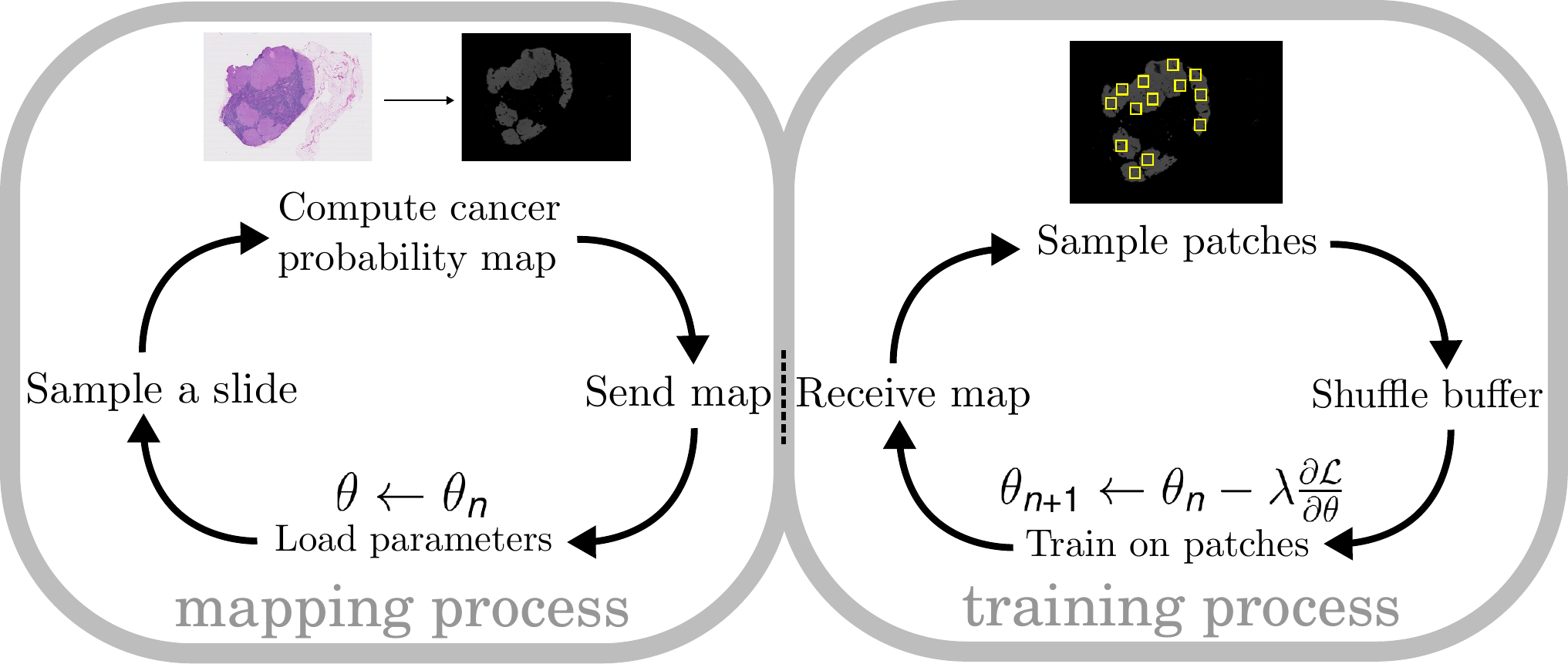}
  \caption{Simplified representation of the training pipeline: two processes run synchronously and in parallel. A mapping process computes cancer probability maps on slides that are used by a training process to extract patches used to train the model.}
  \label{fig:pipeline}
\end{figure}

The training is performed using two processes. The \textit{mapping process} computes the probability maps on slides using up-to-date model weights. The use of fully convolutional segmentation models for $M_\theta$ allows to speedup the computation of the probability maps by allowing to infer the model on larger patches. The \textit{training process} samples patches from the latest probability map to feed a shuffle buffer from which training batches are built and used to update the model's parameters. The two processes are running synchronously in order to optimize the up-to-dateness of the probability maps from which the patches are sampled; and in parallel allowing the training to run without bottleneck. See the figure \ref{fig:pipeline} for an illustration of the training pipeline.

\paragraph{Slides sampling strategy} The \textit{mapping process} samples malign and benign slides with the same rate. Benign slides are sampled with an emphasis on those containing high segmentation error according to the following probability: 
$$P(S=s^*| \textit{benign}) = \frac{\sum_{i,j} Q_{i,j,1} |s^*, \textit{benign}}{\sum_{s,i,j} Q_{i,j,1}|s, \textit{benign}}$$

The probability maps $Q$ are taken from the latest map computed by the mapping process. A default value is used for slides that haven't been computed yet. We ensure that every slides are sampled at least once every $N$ epochs.

\section{Experiments}

\paragraph{Dataset} We trained a model on the CAMELYON 16 dataset \cite{cam16}, which consists of 400 WSIs. The dataset has been developed for the task consisting of detection of metastases in H\&E stained WSIs of lymph node sections. The tumors have been annotated at the pixel level by pathologists. For training we only used the slide level cancer presence information to train the model in a weakly supervised setup. We used the pixel level annotations for evaluation.

\paragraph{Segmentation model}
 We instantiated $M_\theta$ with a modified U-net\cite{ronneberger2015} network. We replaced the convolutions with separable convolutions. We added residual skip connections and batch-normalization. We set the starting number of filters at 32 in the first layer. The model is trained on a resolution of 1 micrometer per pixel. 

\paragraph{Pre-training}

We pre-trained $M_\theta$ by initializing it with an auto-encoder trained with the same network architecture but without the lateral skip connections. It has been trained to reconstruct the input using the $L_2$ loss.

\paragraph{Data augmentation}

We augmented the patches with random rotation, mirroring, elastic deformation and color jittering.

\paragraph{Hyperparameters} We have set $\beta_0 = 0$, $\beta_1=1$, $\eta=50\%$ and $\alpha=2$. 

\paragraph{Results}
We measured the ability of the model to discriminate malignant and benign slides. We scored the slides with their maximum cancer pixel probability value. Given such scoring, the model could reach a classification performance of $0.89$ ROC AUC.

\begin{figure}
  \centering
        \includegraphics[width=0.328\textwidth,scale=1.0]{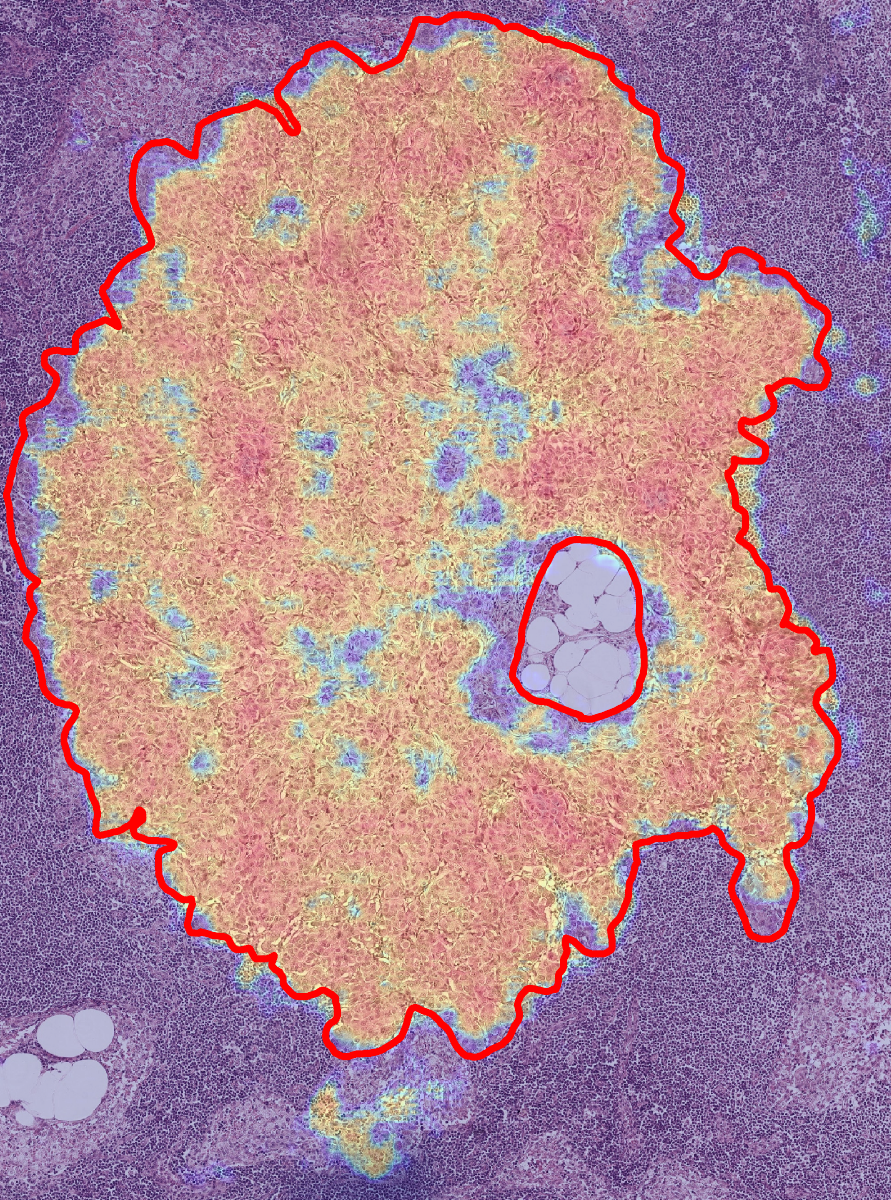}
       \includegraphics[width=0.49\textwidth,scale=1.0]{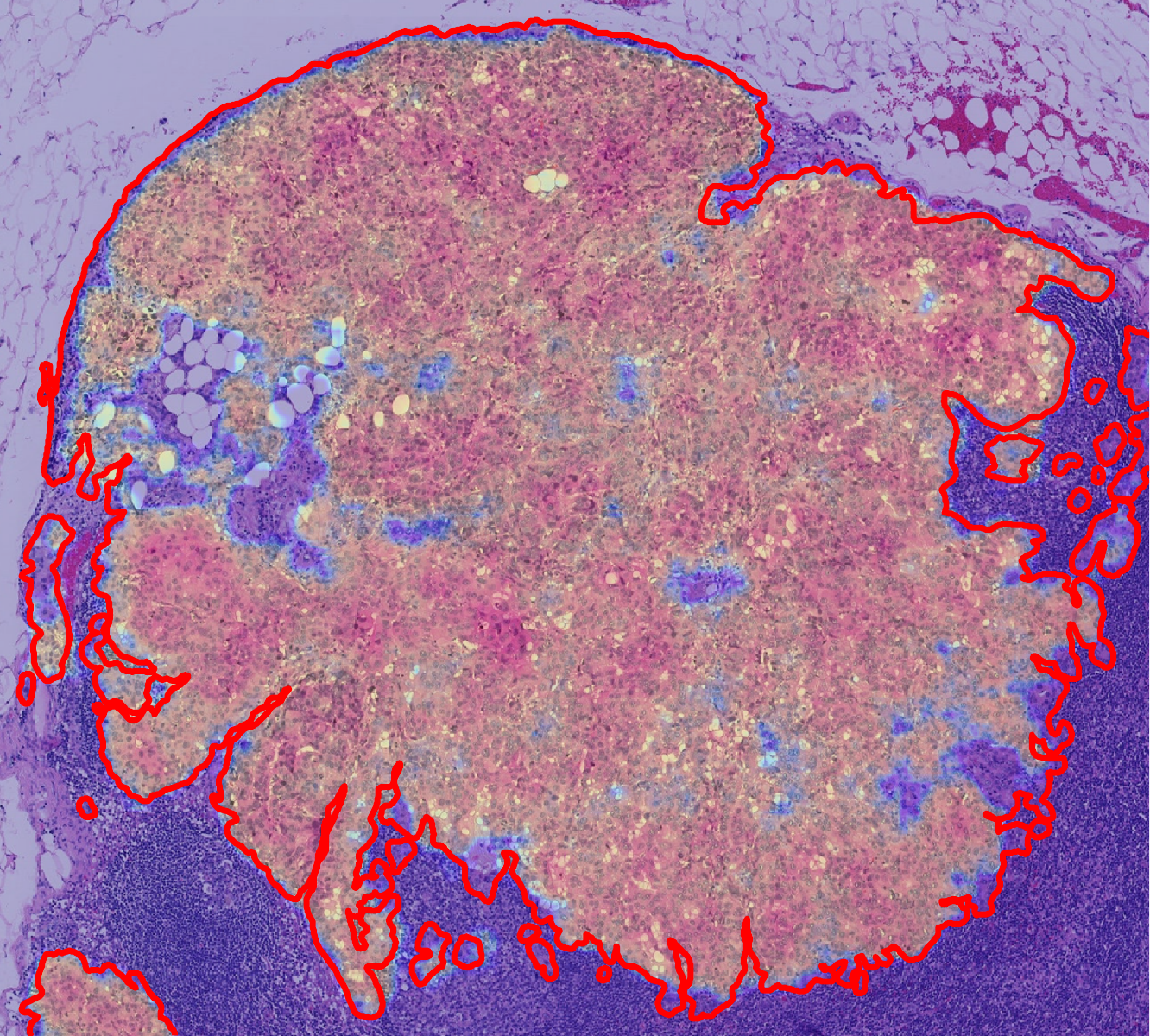}
  \caption{Examples of successfully detected metastases from the test set (image 27 and 1). The cancer probabilities are overlaid on the tissue with a coloring going from blue to red. Red indicates higher probability value. The red outline represents the ground truth annotation.}
  \label{fig:success_seg}
\end{figure}

The quality of the segmentation has been measured using the same metric as the one proposed for the CAMELYON 16 challenge. It is the average sensitivity at 6 predefined false positive rates: $1/4$, $1/2$, $1$, $2$, $4$, and $8$ FPs per whole slide image. The model reached an average sensitivity of $0.43$. The figure \ref{fig:success_seg} shows examples of successfully detected metastases. We can see that the segmentations are consistent with the morphology and have strong agreement with the ground truth. 

\begin{figure}
  \centering
        \includegraphics[width=0.45\textwidth,scale=1.3]{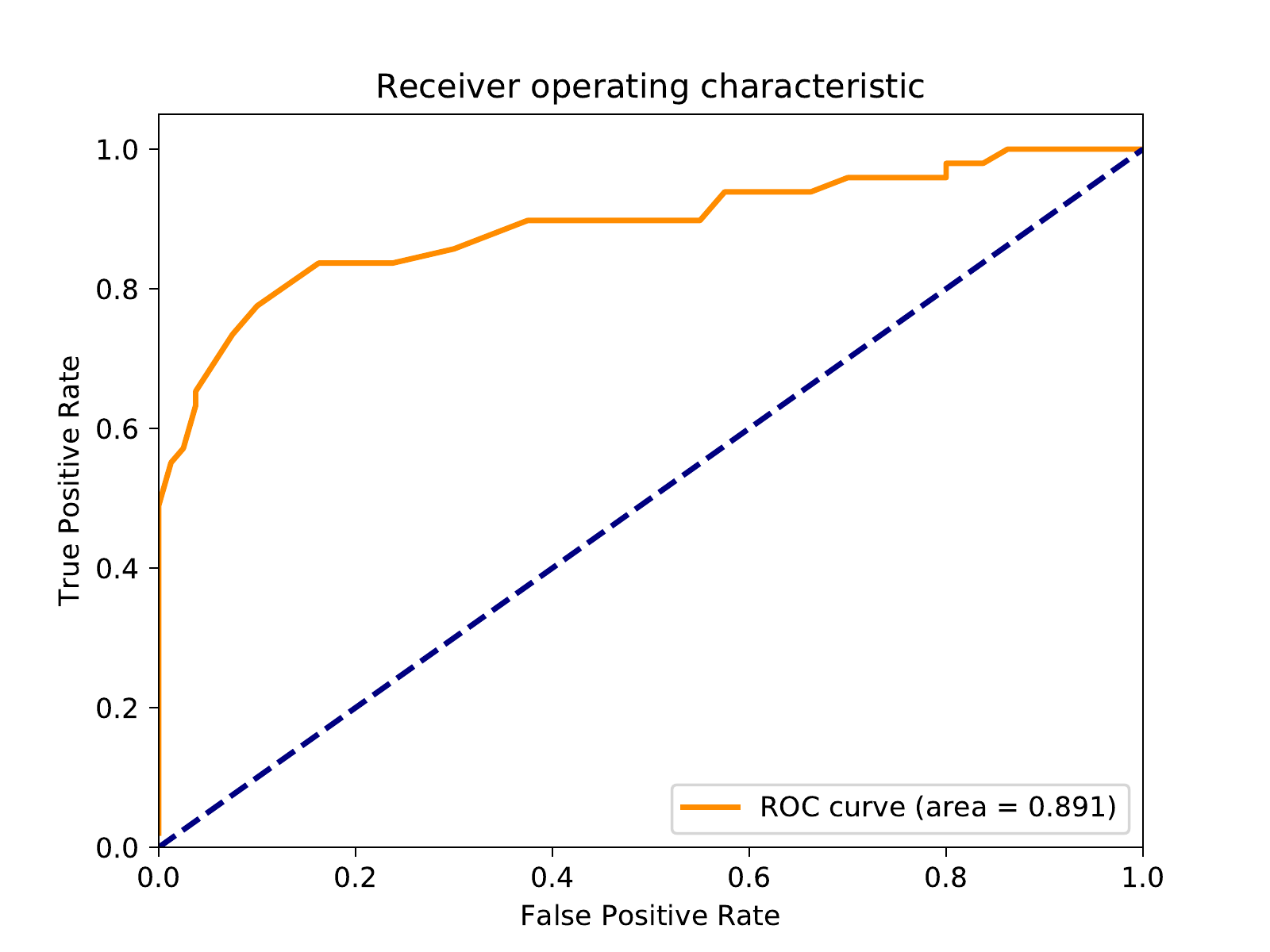}
       \includegraphics[width=0.45\textwidth,scale=1.3]{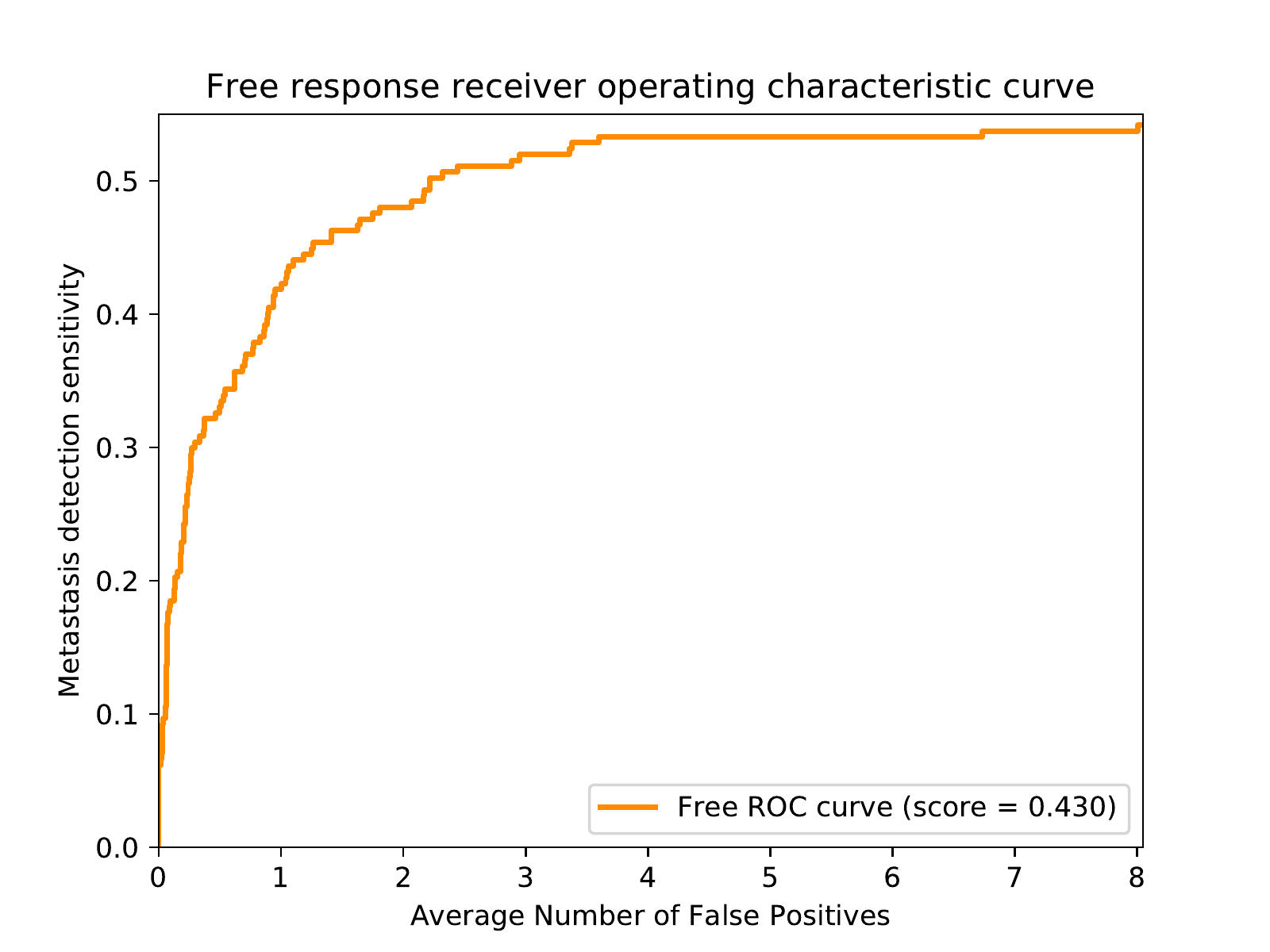}
  \caption{\textbf{left:} Model's ROC curve for slide classification task. \textbf{right:} Model's FROC curve for slide segmentation task.}
  \label{fig:curves}
\end{figure}

\section{Conclusion}

We have proposed a novel method to train a segmentation model at pixel resolution in a weakly supervised setup. We have shown that the model trained with scarce WSI level supervision was able to retrieve cancer localization at pixel resolution. Preliminary experiment have shown promising results in term of classification and segmentation performances inviting for further investigation.
\clearpage
\bibliographystyle{unsrt}  
\bibliography{references}  

\begin{thebibliography}{10}

\bibitem{cam16}
Babak Ehteshami~Bejnordi, Mitko Veta, Paul Johannes~van Diest, Bram van
  Ginneken, Nico Karssemeijer, Geert Litjens, Jeroen A. W.~M. van~der Laak, ,
  and the CAMELYON16~Consortium.
\newblock {Diagnostic Assessment of Deep Learning Algorithms for Detection of
  Lymph Node Metastases in Women With Breast CancerMachine Learning Detection
  of Breast Cancer Lymph Node MetastasesMachine Learning Detection of Breast
  Cancer Lymph Node Metastases}.
\newblock {\em JAMA}, 318(22):2199--2210, 12 2017.

\bibitem{a}
Deepak Pathak, Evan Shelhamer, Jonathan Long, and Trevor Darrell.
\newblock Fully convolutional multi-class multiple instance learning, 2014.

\bibitem{1411.6228}
Pedro~O. Pinheiro and Ronan Collobert.
\newblock From image-level to pixel-level labeling with convolutional networks,
  2014.

\bibitem{1803.07703}
Li~Yao, Jordan Prosky, Eric Poblenz, Ben Covington, and Kevin Lyman.
\newblock Weakly supervised medical diagnosis and localization from multiple
  resolutions, 2018.

\bibitem{1805.06983}
Gabriele Campanella, Vitor Werneck~Krauss Silva, and Thomas~J. Fuchs.
\newblock Terabyte-scale deep multiple instance learning for classification and
  localization in pathology, 2018.

\bibitem{1802.02212}
Pierre Courtiol, Eric~W. Tramel, Marc Sanselme, and Gilles Wainrib.
\newblock Classification and disease localization in histopathology using only
  global labels: A weakly-supervised approach, 2018.

\bibitem{1701.00794}
Zhipeng Jia, Xingyi Huang, Eric I-Chao Chang, and Yan Xu.
\newblock Constrained deep weak supervision for histopathology image
  segmentation.
\newblock 2017.

\bibitem{imagenet_cvpr09}
J.~Deng, W.~Dong, R.~Socher, L.-J. Li, K.~Li, and L.~Fei-Fei.
\newblock {ImageNet: A Large-Scale Hierarchical Image Database}.
\newblock In {\em CVPR09}, 2009.

\bibitem{streamsgd}
Geert~Litjens Hans Pinckaers~M.D., Wouter~Bulten.
\newblock High resolution whole prostate biopsy classification using streaming
  stochastic gradient descent, 2019.

\bibitem{1804.05712}
Hans Pinckaers and Geert Litjens.
\newblock Training convolutional neural networks with megapixel images, 2018.

\bibitem{Momeni438341}
Alexandre Momeni, Marc Thibault, and Olivier Gevaert.
\newblock Deep recurrent attention models for histopathological image analysis.
\newblock {\em bioRxiv}, 2018.

\bibitem{ronneberger2015}
Olaf Ronneberger, Philipp Fischer, and Thomas Brox.
\newblock {U}-{N}et: {C}onvolutional {N}etworks for {B}iomedical {I}mage
  {S}egmentation.
\newblock {\em CoRR}, abs/1505.04597, 2015.

\bibitem{deeplabv3plus2018}
Liang-Chieh Chen, Yukun Zhu, George Papandreou, Florian Schroff, and Hartwig
  Adam.
\newblock Encoder-decoder with atrous separable convolution for semantic image
  segmentation.
\newblock In {\em ECCV}, 2018.

\bibitem{focal_loss}
Tsung-Yi Lin, Priya Goyal, Ross Girshick, Kaiming He, and Piotr Dollár.
\newblock Focal loss for dense object detection, 2017.

\end{thebibliography}

\end{document}